\documentclass[british,english,useAMS,usenatbib]{mn2e}
\usepackage[latin9]{inputenc}
\usepackage{amsmath}
\usepackage{amssymb}
\usepackage{graphicx}
\usepackage{float}
\usepackage{subfig}

\makeatletter

\usepackage{epsfig}
\bibstyle{mn2e}

\usepackage{aas_macros}
\newcommand{\beq}{\begin{equation}}
\newcommand{\eeq}{\end{equation}}

\newdimen\hssize
\newdimen\hdsize 
\title{The environmental dependence of neutral hydrogen in the \textsc{gimic} simulations}
\title[Environmental dependence of HI]
{The environmental dependence of neutral hydrogen in the \textsc{gimic} simulations}

\author[Cunnama et~al.]
{\parbox{\textwidth}{D.~Cunnama,$^{1,2,3}$\thanks{E-mail: \texttt{dcunnama@uwc.ac.za}}
S.~Andrianomena,$^{4}$
C.M.~Cress,$^{1,2}$
A.~Faltenbacher,$^{5}$ 
B.K.~Gibson,$^{6,7,8}$ 
T.~Theuns$^{9,10}$}\vspace{0.4cm}\\
\parbox{\textwidth}{$^{1}$Physics Department, University of the Western Cape, Cape Town 7535, South Africa\\
$^{2}$Centre for High Performance Computing, CSIR, 15 Lower Hope St., Rosebank, Cape Town, South Africa\\
$^{3}$South African Astronomical Observatory, PO Box 9, Observatory 7935, Cape Town, South Africa\\
$^{4}$Centre for Astrophysics, Cosmology \& Gravitation,
and, Department of Mathematics \& Applied Mathematics,\\
University of Cape Town, Cape Town 7701, South Africa\\
$^{5}$School of Physics, University of the Witwatersrand, Private Bag 3, Wits 2050, Johannesburg, South Africa\\
$^{6}$Department of Astronomy \& Physics, Saint Mary's University, Halifax,
Nova Scotia, B3H~3C3, Canada\\
$^{7}$Monash Centre for Astrophysics, School of Mathematical Sciences,
Monash University, Clayton, VIC, 3800, Australia\\
$^{8}$Jeremiah Horrocks Institute, University of Central Lancashire, Preston
PR1~2HE\\
$^{9}$Institute for Computational Cosmology, Department of Physics, University of Durham, South Road, Durham DH1 3LE\\
$^{10}$Department of Physics, University of Antwerp, Campus Groenenborger, Groenenborgerlaan 171, B-2020 Antwerp, Belgium}}

\makeatother
\usepackage{babel}

\date{Accepted for publication in MNRAS. \vspace{-0.6cm}}
\begin{document}

\maketitle
\pagerange{\pageref{firstpage}--\pageref{lastpage}} \pubyear{2013}

\label{firstpage}
\begin{abstract}
We use the Galaxies-Intergalactic Medium Interaction Calculation (\textsc{gimic}) cosmological hydrodynamic simulation at $z=0$ to study the distribution and environmental dependence of neutral hydrogen (\textsc{Hi}) gas in the outskirts of simulated galaxies. This gas can currently be probed directly in, for example, Ly$\alpha$ absorption via the observation of background quasars. Radio facilities, such as the Square Kilometre Array, will provide a complementary probe of the diffuse \textsc{Hi} in emission and will constrain the physics underpinning the complex interplay between accretion and feedback mechanisms which affect the intergalactic medium. We extract a sample of 488 galaxies from a re-simulation of the average cosmic density \textsc{GIMIC} region. We estimate the neutral hydrogen content of these galaxies and the surrounding intergalactic medium within which they reside. 
We investigate the average \textsc{Hi} radial profiles by stacking the individual profiles according to both mass and environment.
We find high \textsc{Hi} column densities at large impact parameters in group environments and markedly lower \textsc{Hi} densities
for non-group galaxies. We suggest that these results likely arise from the combined effects of ram pressure stripping and tidal interactions present in group environments.
\end{abstract}


\section{Introduction}
\label{sec:intro} 

With the advent of new radio facilities, including MeerKAT\footnote{\tt http://www.ska.ac.za/meerkat/\rm}, ASKAP\footnote{\tt http://www.atnf.csiro.au/projects/askap\rm} and the SKA\footnote{\tt http://www.skatelescope.org/\rm}, there is growing interest in understanding the distribution and characteristics of the fundamental baryonic 
building-block of stars and galaxies - neutral hydrogen (\textsc{Hi}) - both in and around these galaxies. These facilities, and
the surveys they will perform, will generate an abundance of data relating to, amongst other things, the spatial distribution of \textsc{Hi}, its 
kinematics, and its physical state.  The physical extent and structure of low-column density \textsc{Hi} gas in the disks and, especially, halos of galaxies are a powerful probe of the efficiency with which energy from supernovae, massive star radiation and galactic winds couple to the surrounding interstellar medium and dilute halo gas \citep[e.g.][]{Pilkington-11, Stinson-12}.

Surveys such as THINGS \citep{Walter-08}, the WRST HI filament survey \citep{Popping-11} and the VLA Imaging of Virgo Spirals in Atomic Gas project \citep{Chung-09} have provided detailed views of the dynamics of \textsc{Hi} in emission in nearby galaxies, however with current instruments it is still very difficult to probe \textsc{Hi} column densities below $\sim$10$^{17}$~cm$^{-2}$; this acts to restrict our knowledge of the state and distribution of cold gas in the outskirts of galaxies, including the halo, and out to the virial radius.  As noted before, knowledge of this gas constrains directly the interplay between infall and galactic winds in galaxy formation. There have been attempts at deep and/or spatially comprehensive observations of nearby galaxies \citep[e.g.][]{Chynoweth-08, Pisano-11}, in order to characterise the low column density gas at large impact parameters, but these observations are severely limited by the sensitivity of current radio telescopes, with detection limits of order $\
sim$10$^{17-18}$~cm$^{-2}$ \citep{Oosterloo-07}.

Probing such low column density halo gas ($<$ 10$^{17}$~cm$^{-2}$) can be approached in a complementary manner using absorption features from foreground gas associated with low redshift galaxies seen in the spectra of background quasars. Such lines include indirect proxies for cold gas, including \textsc{M}g\textsc{II} \citep{Bordoloi-11}, and direct observations of Ly$\alpha$ at high impact parameters and at redshifts of $z < 0.2$ \citep[e.g.][]{Prochaska-11,Tripp-05, Tumlinson-13}.
Through observational campaigns such as these, detections of \textsc{Hi} at distances of 50-250 kpc from the host galaxy have yielded important clues as to the nature of the gas in these regions and hinted at the possibility of detecting large reservoirs of cold gas residing at these 
galacto-centric radii. 

Of particular interest, is the detection of an environmental dependence in the amount of \textsc{MgII} absorption at high impact parameters presented in \citet{Bordoloi-11}.  These authors present a simple model in which the radial distribution of cold halo gas for (all) galaxies
is independent of environment. According to this model the observed environmental dependency is simply an apparent one in which the larger extent of gas profiles in group galaxies is purely a superpositional effect (in condensing an inherently three-dimensional group in redshift-space to a
two-dimensional image plane). While \textsc{MgII} absorption cannot be used as a direct proxy for \textsc{Hi} it has long been used as a tracer for cold gas and the environmental dependence observed in \textsc{MgII} may be observable in \textsc{Hi} absorption.

Since three quarters of the baryonic material in the Universe is hydrogen, when developing a complete theory of galaxy formation it is important
to be able to reproduce the distribution and phases of this hydrogen gas. Several studies have succeeded in modelling accurately the \textsc{Hi}
column density distribution at $z$=3 \citep[e.g.][]{Pontzen-08, Razoumov-08, Tescari-09, Altay-11, McQuinn-11}, but most relevant to our investigation are efforts to model \textsc{Hi} at low redshift, including those of \citet{Popping-09} and \citet{Duffy-11}.
\citet{Popping-09} reproduced the low-redshift \textsc{Hi} mass function using a simple pressure-based prescription for calculating
the neutral fraction. \citet{Duffy-11} expanded upon this by implementing a multiphase treatment for calculating the various states of hydrogen; the latter also claim that the \textsc{Hi} mass function is subject to only weak evolution and is insensitive to whether AGN feedback is included, a claim supported by \citet{Dave-13}. It is important to note that these studies have ignored the effect of local sources, recently \citet{Rahmati-13b} did include the effect of local sources and showed that at $z=0$ it is important for $N_{HI} > 10^{21}$ and at $z=3$ for $N_{HI} > 10^{17}$. In addition \citet{Rahmati-13a} calculated $z=0$ HI distribution predicted by OWLS and compared it to observations, finding good agreement.\\

The simulation we employ in this study is the mean density region (the so-called `0$\sigma$' region)
of the \textsc{gimic} suite of simulations; these have been shown to successfully reproduce many observables, including the satellite luminosity
function, stellar surface brightness distributions, and the radial distribution of metals \citep[e.g.][]{Crain-09, Font-11, McCarthy-12}.
The \textsc{gimic} simulations have employed very efficient supernova feedback, which itself was required to obtain an empirically-supported low star formation efficiency \citep[e.g.][]{Crain-09, Schaye-10} and reproduce accurately the enrichment of the intergalactic medium
\citep[e.g.][]{Aguirre-01, Oppenheimer-06, Wiersma-11}. In addition, the number density of galaxy--absorber groups in the \textsc{gimic} simulations
have been shown to be consistent with observations \citep{Crighton-10}.\\

We apply the prescription outlined by \citet{Duffy-11}, to estimate the neutral fraction for each of the SPH particles in our simulation. We then investigate the radial distribution of the neutral hydrogen around the galaxies, in order to compare to absorption observations
and, in particular, we investigate the effect of the galactic environment on this radial distribution. We have extracted a catalogue of almost
500 galaxies, spanning a wide range in mass and calculated their corresponding neutral hydrogen content and distribution within $\sim$250~kpc of
their respective galactic centres. Using this catalogue, we have created radial neutral hydrogen column density profiles for each galaxy and stacked these profiles according to the galactic environment in which they reside,
in attempt to investigate putative environmental effects.\\

Our results show a marked difference in the extent of the radial neutral hydrogen column density profiles when comparing both group and non-group galaxies as well as central and satellite galaxies. We attribute this difference to the ram pressure experienced primarily by 
satellite galaxies in the group environment which results in cool gas being stripped from the cores and redistributed to larger radii. In addition we demonstrate that the group and satellite galaxies experience greater ram pressure forces and have correspondingly lower neutral hydrogen fractions.\\

This paper is arranged as follows: in Section \ref{secsim}, we present our simulation and its associated galaxy catalogue. We next emphasise the
methodology by which we estimate the neutral fraction for the gas particles, in Section \ref{secextraction}. Section \ref{sec:HI-Maps} outlines the means by which we (a) create maps of the \textsc{Hi} column density for each of our galaxies and (b) extract and stack the radial \textsc{Hi} 
column density profiles. Our results are presented in Sections \ref{sec:HI-Maps}, \ref{sec:rampressure} and \ref{sec:Superposition}. Our conclusions are then presented in Section \ref{secconclusion}.

\begin{table*}
\begin{center}
\caption{The simulated sample of galaxies and their environments}
\begin{tabular}{|l|c|c|c|} \hline Stellar Mass & Number of & Number of  & Number
of \\
$[h^{-1} {\rm M}_\odot]$ & Galaxies & Group Galaxies  & Field Galaxies \\ \hline $10^{9}
- 10^{10}$ & 282 & 110  & 172 \\ $10^{10} - 10^{11}$ & 155 & 88  & 67 \\ $>
10^{11}$ & 51 & 34  & 17 \\ \hline 
\end{tabular}
\end{center}
\end{table*}

%
%
%

\section{Simulation and Galaxy Catalogue}
\label{secsim}

The simulation we have utilised is one from the \textsc{gimic} suite of simulations which are described in detail
by \citet{Crain-09}. The \textsc{gimic} simulations are fully cosmological and hydrodynamical, and were designed to 
investigate the interaction between galaxies and the intergalactic medium. The simulations were run
using \textsc{Gadget-3}, an updated version of the publically available gravitational N-body+SPH code \textsc{Gadget-2} \citep{Springel-05}.

The complete suite of simulations consists of resimulations of five nearly spherical regions of $\sim$20~Mpc in radius which
were extracted from the Millennium Simulation \citep{Springel-05a}. The regions were picked to have varying overdensities at $z$=1.5 of ($+$2,$+$1,0,$-$1,$-$2)$\sigma$,
where $\sigma$ is the root-mean-square deviation from the mean density on this scale. For our work here, we use only the high resolution
0$\sigma$ region, to generate our base galaxy catalogue.

The cosmological parameters adopted are the same as those for the Millennium Simulation and correspond to a $\Lambda$CDM model with $\Omega_{m}=0.25$,
$\Omega_{\Lambda}=0.75$, $\Omega_{b}=0.045$, $\sigma_{8}=0.9$, $H_{0}=100h^{-1}$~km~s$^{-1}$~Mpc$^{-1}$, $h=0.73$,
$n_{s}=1$ (where $n_{s}$ is the spectral index of the primordial power spectrum). The value of $\sigma_{8}$
is roughly 2-sigma higher than that inferred from recent CMB data \citep{Komatsu-11}, which will affect the relative numbers of 
Milky Way-scale systems, but should not impact upon their individual, internal, characteristics.

The high resolution 0$\sigma$ \textsc{gimic} region was evolved from $z=127$ down to $z=0$ using \textsc{Gadget-3}. The radiative cooling rates are computed on an elemental basis by interpolating pre-computed \textsc{CLOUDY}  \citep{Ferland-98} tables containing cooling rates as a function of density, temperature, and redshift \citet{Wiersma-09}. The cooling rates take into account the presence of the cosmic microwave background and photoionisation from a \citet{Haardt-96} ionising UV background. The background is switched on at $z=9$ where the entire volume is rapidly ionised. Star formation is tracked following the prescription of \citet{Schaye-08} which, by construction, reproduces the observed Kennicutt-Schmidt relation \citep{Kennicutt-98}. Individual chemical elements are re-distributed through a timed release by both massive and intermediate mass stars,
as described by \citet{Wiersma-09}.

Feedback is implemented using the kinetic model of \citet{Dalla-08} with the initial wind velocity set at 600~km/s and a mass loading parameter $\eta$=4. This choice of parameters results in a good match to the peak cosmic star formation rate \citep{Crain-09} and reproduces a number of X-ray and optical scaling relations for normal disc galaxies \citep{Crain-10}. As shown by \citet{McCarthy-11} and \citet{McCarthy-12}, 
the \textsc{gimic} simulations also produce realistic spheroidal components around $\sim$L$^\star$ galaxies. The \textsc{GIMIC} simulations do suffer from some overcooling for large galaxies with stellar mass much greater than $10^{10} M_{\odot}$ \citep{Crain-09,McCarthy-12} and the same is true for the intragroup medium which could have consequences for ram pressure stripping.

\subsection{Sample Selection}
\label{subsec:sampleselection}

In this study we focus on the $z=0$ snapshot from the high-resolution $0\sigma$ \textsc{gimic} Simulation. From this we identify galaxies using \textsc{subfind} \citep{Springel-01,Dolag-09} and select only those galaxies with a stellar mass above $10^{9}h^{-1}$~M$_\odot$. We then use these galaxies to infer the associated environment information for each system
i.e., whether each galaxy resides in a cluster, group, or the field.
This is done by assigning each galaxy as a group or non-group galaxy, dependent upon the average distance of the 7 nearest neighbours \citep{Faltenbacher-10}. If the average distance is less than 1 Mpc then the galaxy is associated with a group whereas if the average distance is greater than 2 Mpc we classify the galaxy as a non-group galaxy. In this way we ensure a clear distinction between group and field galaxies for our sample. Our sample is presented in Table~1.

In addition we separate our sample into central and satellite galaxies, with a central galaxy being defined as the largest galaxy in its parent FOF halo.

We are particularly interested in the gas residing in the outskirts of the galaxies. We extract a cubic region of $600 h^{-1}$ kpc on a
side around each galaxy from the original snapshot. As we are investigating the properties of the extended \textsc{Hi} around galaxies we do not limit ourselves to gravitationally bound particles; instead, we consider all particles in the environment of the galaxy, including all substructure
such as dwarf galaxies and high-velocity clouds. This method of selection will result in the inclusion of some gas from neighbouring and satellite galaxies, the consequences of which we discuss in section \ref{sec:Superposition}.

\section{Gas and Neutral Hydrogen Extraction}
\label{secextraction}

To perform our analysis, we need to first estimate the fraction of neutral hydrogen associated with the gas within our simulation. In order to do this, we follow the prescription presented by \citet{Duffy-11}. In this method, the assumption is made that the ionising UV radiation originates
from purely extragalactic sources and that the ionising effect of internal stellar sources is insignificant. This is a reasonable assumption since the gas in the centre of a galaxy can be considered as largely self-shielded and the gas density in the outer regions is low enough to render
the effects of collisional ionisation negligible. As noted by \citet{Popping-09}, these assumptions are also consistent with the findings of \citet{Dove-94} who found that the \textsc{Hi} column density is more sensitive to external radiation than to that of the host galaxy.

In order to calculate the neutral fraction, the hydrogen mass fraction of each particle is first taken to be 0.75. We then calculate the fraction of this hydrogen which can be considered neutral. This is done by sub-dividing the hydrogen into one of the three phases:

\paragraph{Interstellar Medium}

As gas in the interstellar medium (ISM) cools and its density increases it can no longer be reliably modelled in our simulations, this gas is associated with the star-forming interstellar medium in galaxies. If a particle has exceeded the critical particle density $n_{H}$=0.1~cm$^{-3}$ and has a temperature less than 10$^5$~K it is assumed to be star-forming and is modelled according to an effective polytropic Equation-of-State \citet{Schaye-08}.

This ``EoS'' gas, or interstellar gas, lies in very dense regions where self-shielding is important, and we assume that the ISM exists as either H$_2$ or \textsc{Hi}. To separate the two phases, we utilise the empirical ratio relating H$_2$ and \textsc{Hi} surface densities to the local ISM pressure measured by THINGS \citep{Leroy-09}:
\begin{equation}
 R_{surf} = \left(\frac{P/k}{10^{4.23} K cm^{-3}} \right)^{0.8}.
\end{equation}
We then use this ratio to compute the mass of interstellar gas in neutral hydrogen.

\paragraph{Self-shielded Gas}

If the ISM is not ``star forming'' we approximate the onset of self-shielding using a pressure prescription similar to \citet{Popping-09}. We use
the same pressure threshold presented by \citet{Duffy-11} of $P_{shield}/k\sim 150$~K~cm$^{-3}$, which has been shown to reproduce the cosmic \textsc{Hi} 
density in the mass range $10^{10}-10^{11}h^{-1}$~M$_\odot$ by the ALFALFA Survey \citep{Martin-10}, albeit with slightly different feedback to the \textsc{gimic} simulations.
Gas that exceeds $P_{shield}$ and has a temperature less than $10^{4.5}$~K is assumed to be fully self-shielded and therefore fully neutral.


\paragraph{Optically-thick Gas}

For gas that lies at intermediate densities ($10^{-4}-10^{-1}$~cm$^{-3}$), we can no longer assume the ISM is optically-thin and exposed to a
uniform background radiation field. For this gas, we calculate the neutral fraction according to the analytic description presented by
\citet{Popping-09}. Essentially, the degree of ionisation is determined by the balance between the photo-ionisation rate and the recombination
rate of the gas. The photo-ionisation rate we use is given by the CUBA model of \citet{Haardt-01} - i.e., $\Gamma_{HI}\approx10^{-13}$~s$^{-1}$ at redshift $z$$\sim$0.

The recombination rate is calculated using the analytical function of \citet{Verner-96}, over the temperature range 3K to 10$^{10}$K. This analytical approach differs slightly from the method utilised by \citet{Duffy-11}, wherein \textsc{CLOUDY} lookup tables were interpolated, to calculate the 
neutral fraction for the gas.

\section{HI Maps and Radial Profiles}
\label{sec:HI-Maps}

In order to convert our 3D simulated data cubes into 2D \textsc{Hi} column
density maps which can be used to mimic observed \textsc{Hi} maps, we make use of the \tt vista \rm routine within 
\textsc{tipsy}.\footnote{\tt http://www-hpcc.astro.washington.edu/tools/tipsy/tipsy.html\rm}
As we have used a cubic region of $600 h^{-1}$ kpc on a side around each galaxy, the line-of-sight velocity interval will be smaller than that used by observers, while this does make our comparison with observations slightly inconsistent we do not believe it will greatly affect our qualitative results.
From these projected maps, we can analyse the radial column density distribution for each of the galaxies. This is done by taking
the average \textsc{Hi} column density in successive annuli from the centre of mass of each galaxy, out to an impact parameter of $\sim$250~kpc.
Having done so, we then have the radial distribution for each of the galaxies in our sample, allowing us to then stack according to stellar mass and 
environment. 

\subsection{Group versus Non-Group galaxies}
\label{subsec:GroupvsNonGroup}

In order to quantify the difference between group and non-group galaxies we take the mean \textsc{Hi} column density in each radial bin. We divide our 
sample into three stellar mass bins, to ensure that the more extended profiles of the larger galaxies do not overwhelm the signal from the smaller galaxies (without compromising unduly on the statistics of each mass bin). 

\begin{figure}
 \subfloat[Stellar Mass: $10^{9}-10^{10}  h^{-1}$~M$_\odot$]{\includegraphics[scale=0.69]{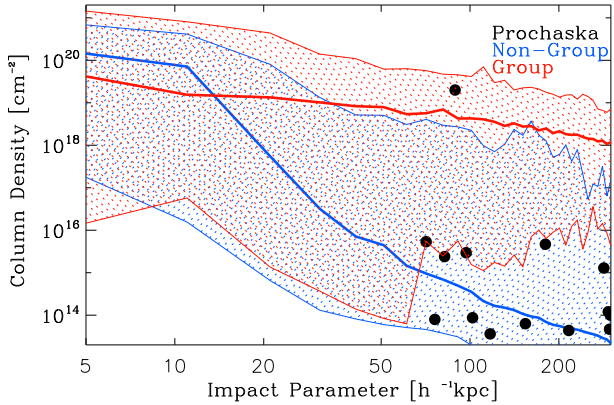} }\\
 \subfloat[Stellar Mass: $10^{10}-10^{11}  h^{-1}$~M$_\odot$]{\includegraphics[scale=0.69]{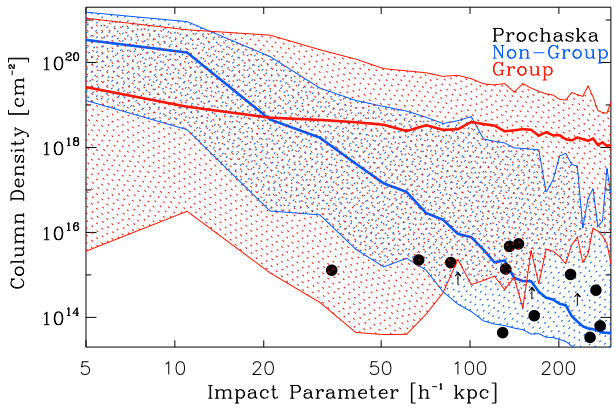} }\\
 \subfloat[Stellar Mass: $>10^{11}  h^{-1}$~M$_\odot$]{\includegraphics[scale=0.69]{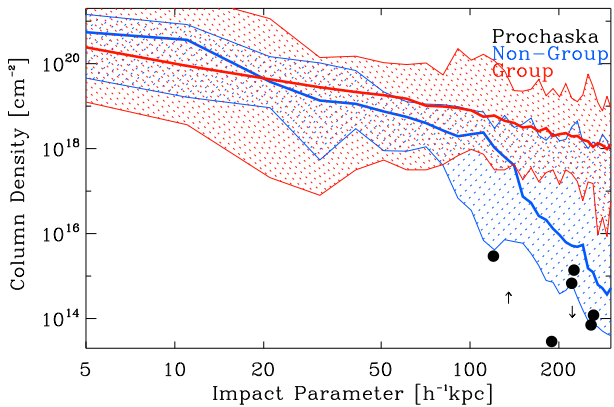}   }

\caption{The solid lines represent the median of the stacked mean \textsc{Hi} column densities as a function of impact parameter for galaxies in the 
stellar mass ranges (a) $10^{9}-10^{10}  h^{-1}$~M$_\odot$ (b) $10^{10}-10^{11}  h^{-1}$~M$_\odot$ (c) $>10^{11}  h^{-1}$~M$_\odot$. Black dots correspond to the observations of background QSO absorbers along the lines-of-sight to foreground (a) $<$0.1L$^\star$  (b) $\sim$0.1$-$1L$^\star$ (c) $\sim$L$^\star$ galaxies, from \citet{Prochaska-11}. Coloured regions include the 5th and 95th percentiles for each bin.\\
In all mass ranges the radial profiles of the group galaxies are noticeably extended when compared to the galaxies in a non-group environment while in general our results are consistent with the observations of \citet{Prochaska-11}.
\label{fig:Stacked-radial-HI-9-10}}

\end{figure}

As can be seen in each of the three mass ranges (Figure ~\ref{fig:Stacked-radial-HI-9-10}), while there exists a large scatter,
there is a large physical difference in the shape of the stacked radial profile of group and non-group galaxies. The non-group galaxies radial
profile drops steadily as might be expected for a standard isolated halo \citep{Koopmann-06}, however the group galaxies show a more extended halo of cool gas,
resulting in mean \textsc{Hi} column densities as large as $\sim$10$^{18}$~cm$^{-2}$ as far out as $\sim 150   h^{-1}$~kpc from the galactic centre. Overplotted in each of the figures are column density measurements from QSO observations for galaxies of similar stellar mass, from \citet{Prochaska-11} and \citet{Tripp-05}. In each of the cases, our radial column densities, particularly for isolated galaxies, correspond fairly well with the observations. It is important to note that we have not imposed/constructed any \it a priori \rm superposition of halos, as was necessarily implemented in the analysis of \citet{Bordoloi-11}.  Each radial profile here is constructed using neutral gas surrounding each individual galaxy, regardless of environment, we conclude that there exists in our simulated galaxies, a physical difference in radial distribution of neutral hydrogen in group environments and in isolated galaxies.

In comparing the radial profiles in each environment, our relatively large sample allows us to observe several systematic trends. For the $10^{9}-10^{10}  h^{-1}$~M$_\odot$ stacked stellar mass galaxies, the difference between those galaxies that reside in groups and those that do not is particularly apparent; we attribute this to the fact that the smaller satellite galaxies are more likely to have undergone ram pressure stripping
on their infall through the group environment \citep{Freeland-11,Hess-13}, resulting in the gas in the stellar disk being removed and being distributed further from
the galactic centre as suggested by \citet{Bahe-13a,Bahe-13b}. Further to this the fact that the 5th percentile (shaded region of Figure \ref{fig:Stacked-radial-HI-9-10}) of the group galaxies drops as low as $\sim$10$^{16}$~cm$^{-2}$ near the centre of the galaxies suggests the possible removal of cold gas from the central parts. Another possible explanation could be that stellar feedback has a larger effect in these smaller galaxies, however the fact that the non-group galaxies have similar stellar masses leads us to believe that this is not the case and that ram pressure stripping is primarily responsible.\\

In comparing the larger mass bins, we see that our radial column densities once again compare well with the few reliable observations that exist,
however the environmental dependence becomes less extreme. We can attribute the trend that larger galaxies exhibit more extended profiles firstly to the fact that the galaxies are physically larger, but more so, larger galaxies with deeper potential wells, are generally more resistant to ram pressure stripping and have undergone fewer infall events than smaller galaxies. Nonetheless, we still observe a difference in the radial profiles of the group galaxies versus the non-group in that the radial profiles of the group galaxies are still noticeably more extended than those of the non-group sample.\\

We note that there does exist some mismatch between our simulated galaxies and the observations of \citet{Prochaska-11} for group galaxies, and in fact the largest observed column density at high impact parameter in Figure \ref{fig:Stacked-radial-HI-9-10}(a) is that of \citet{Tripp-05} which the authors suggest may be the result of the absorber being a dwarf galaxy or high velocity cloud. In this case we are forced to concede that our estimates of the HI column density, while consistent with non-group galaxies, do not seem to match observations for group galaxies at large radii. Deep observations of \textsc{Hi} emission by, for example \citet{Oosterloo-07} and \citet{Fernandez-13} have shown \textsc{Hi} column densities of $\sim$10$^{20}$~cm$^{-2}$ at distances of 20-30 kpc from nearby galaxies, however observations of $>$10$^{18}$~cm$^{-2}$ column density \textsc{Hi} emission at greater than 50 kpc to not exist. Another possible explanation for the mismatch between our group galaxies and current 
observations is that the feedback mechanisms employed in our simulations are distributing the cold gas to too large a radii, or the SPH formalism, which tends to produce disks which are too thick, is making the simulated galaxies prone to ram pressure stripping.\\

Recent work by \citet{Stinson-12} has compared simulations to the same observational datasets using various feedback schemes and conclude that an incorrect implementation of feedback will very quickly result in cool \textsc{Hi} properties which do not match the observations of \citet{Prochaska-11}. We qualitatively compare our results with those of \citet{Stinson-12} comparing profile, amplitude and extent in HI against the non-group profile from our galaxies and find that our feedback scheme as well as our method for calculating neutral content of the cold gas seems to naturally lead to an extended cool gas halo which is consistent with existing observations.

\subsection{Central versus Satellite Galaxies}
\label{subsec:CentralvsSatellite}

As a further test of the environmental differences of the neutral hydrogen distribution in our simulated galaxies we reclassify our galaxy sample according to their position in the parent halo, namely central galaxies and satellite galaxies. Central galaxies are classified as the largest galaxies in their parent FoF halo and Satellite galaxies are then the rest of the galaxies in the halo.\\

\begin{figure}
 
 \subfloat[All Gas]{\includegraphics[scale=0.77]{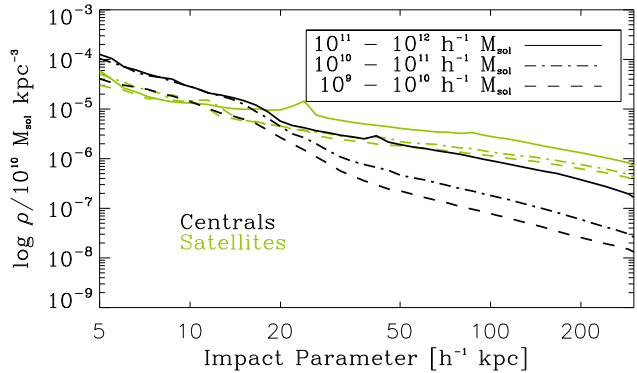}  }\\
 \subfloat[Cold Gas]{\includegraphics[scale=0.77]{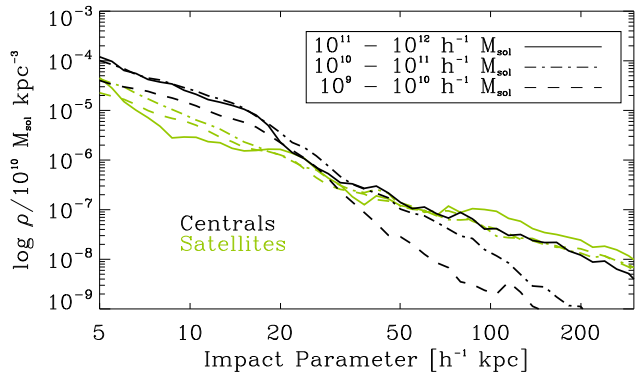} }\\ 
 \subfloat[Neutral Hydrogen ]{\includegraphics[scale=0.69]{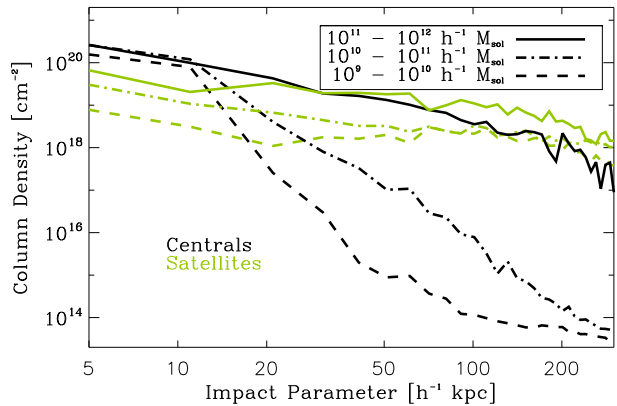}  } 
 \caption{The lines represent the median of the (a) total gas density (b) cold gas density and (c) stacked mean \textsc{Hi} column densities as a function of impact parameter for galaxies in the stellar mass ranges $10^{9}-10^{10}  h^{-1}$~M$_\odot$, $10^{10}-10^{11}  h^{-1}$~M$_\odot$ and $10^{11}-10^{12}  h^{-1}$~M$_\odot$. The radial profiles for satellite galaxies extend, at high column densities, out to large impact parameters whereas the central galaxies drop off sharply with increasing radius.
    \label{fig:StackedCentralvsSat}}
\end{figure}

We perform the same stacking of the radial \textsc{Hi} column density profiles as in Section \ref{subsec:GroupvsNonGroup}, seperating the galaxy sample firstly into central versus satellite galaxies, and secondly by mass. The results are shown in Figure \ref{fig:StackedCentralvsSat} and we can see a clear difference in the extent of the radial profiles of the central and satellite galaxies. The satellite galaxies exhibit large column densities out to large impact parameters while the central galaxies drop off sharply with increasing impact parameter. In addition to the \textsc{Hi} column density profiles we include profiles of the total gas and cold gas($< 2\times 10^4$~K) in order to further illustrate the difference between the satellite and central galaxies. We also observe that satellite galaxies have decreased gas density near the centre with extend profiles at large radii. Figure \ref{fig:StackedCentralvsSat} seems to suggest that the satellite galaxies' neutral gas has been stripped off the host 
galaxy and 
redistributed into the intergalactic medium as it is no longer strictly associated with the satellite galaxy, the mechanism for this stripping can again be attributed to the ram pressure experienced by the infalling satellite galaxies in the group environment.

\section{Ram Pressure stripping}
\label{sec:rampressure}

The previous result suggest that there exists some process by which the cold gas and more precisely the neutral hydrogen in group environments is being dissociated from its host halos and galaxies and being redistributed throughout the intergalactic medium. One possible explanation for this effect is that in the group environment ram pressure stripping is acting  more efficiently on the galaxies than in a field environment. In order to investigate this we compute the ram pressure exerted on individual galaxies.

\subsection{Computing Ram Pressure stripping}

\begin{figure}
  \subfloat[][]{\includegraphics[scale=0.45]{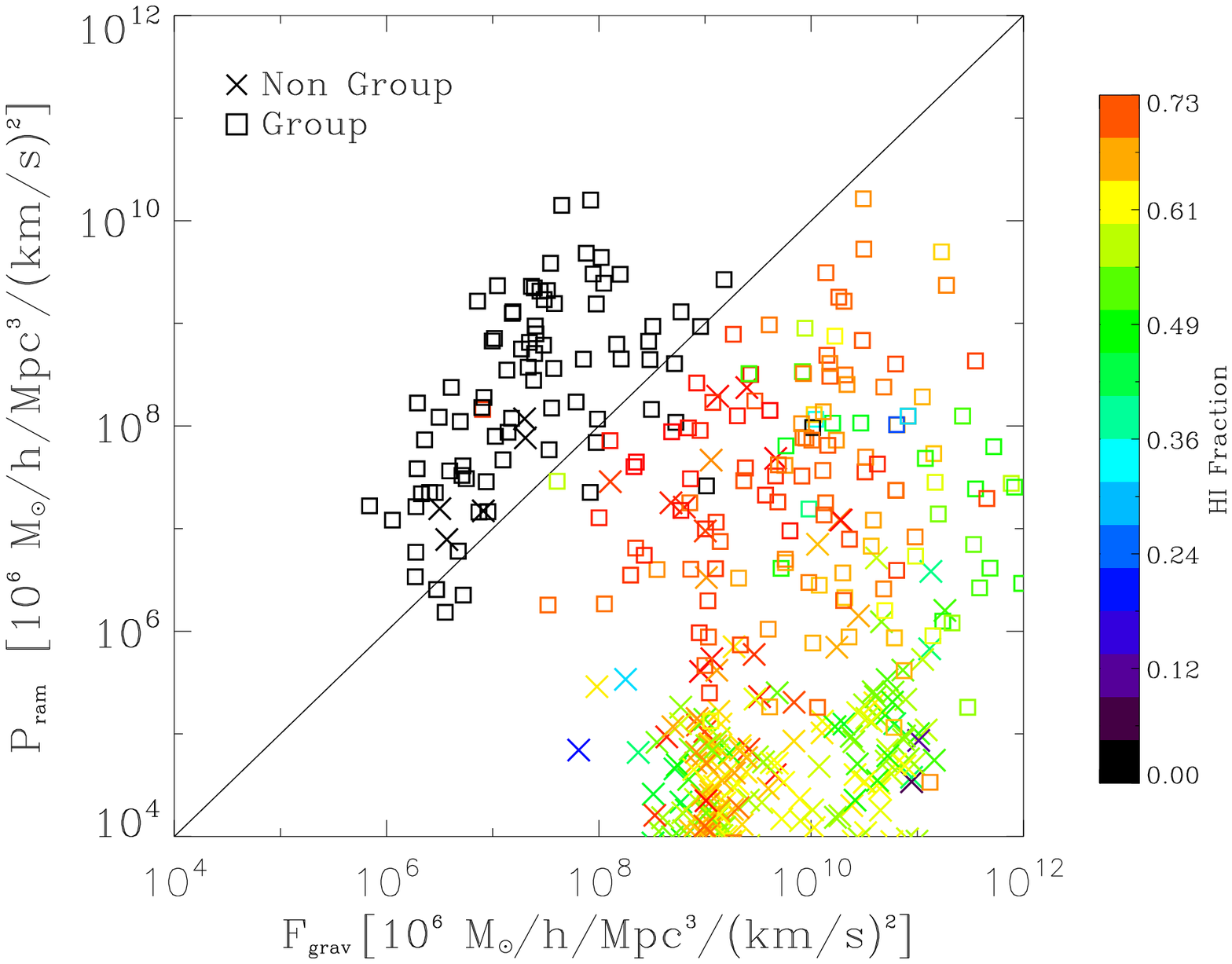}}\\
  \subfloat[][]{\includegraphics[scale=0.45]{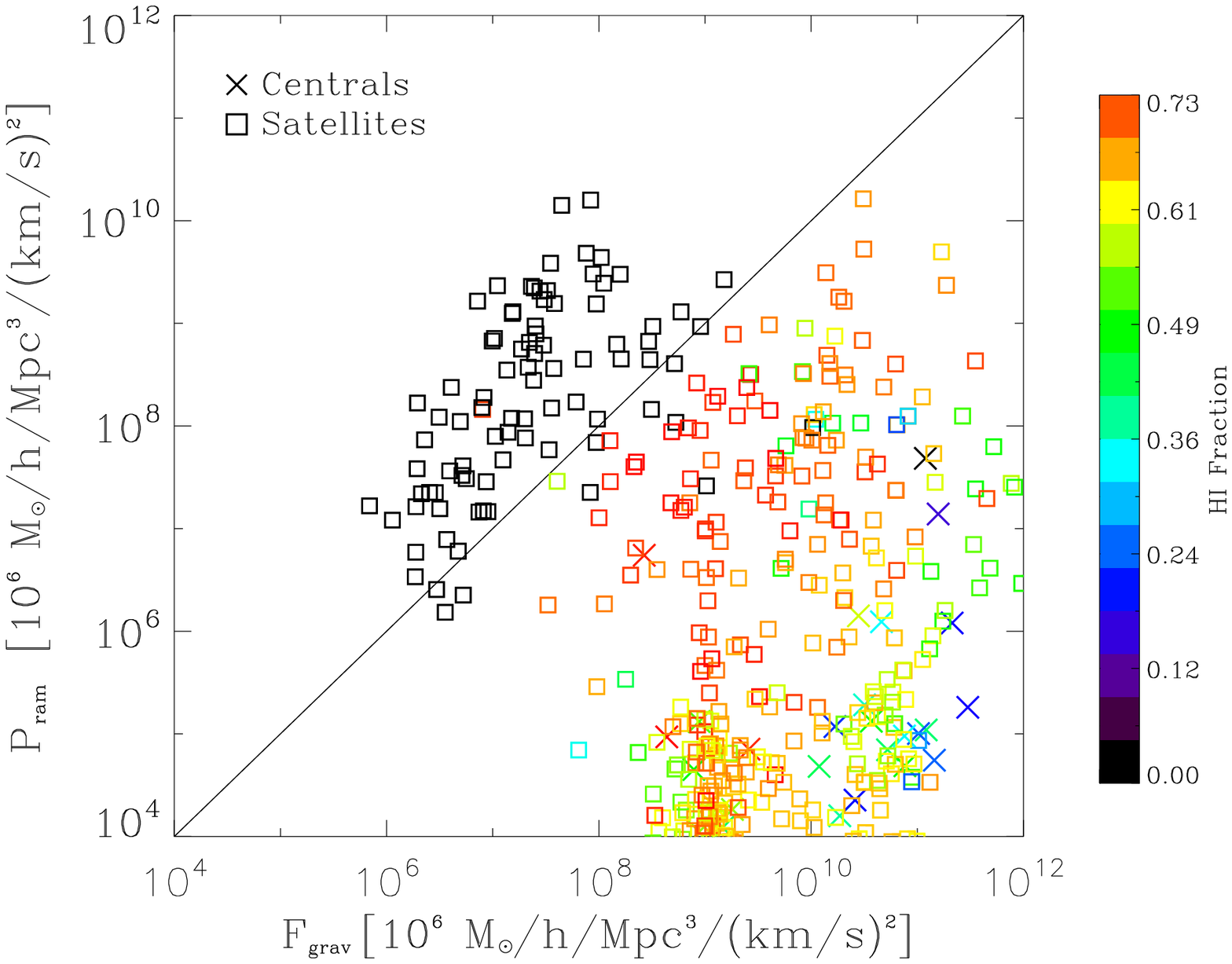}}
  \caption{\label{fig:carthycond}The axes represent the ram pressure exerted on each satellite(vertical-axis) against the gravitational restoring force(horizontal-axis) for both the Non-group galaxies(crosses) and the Group galaxies (squares). The colours from black to red indicate increasing HI ratios within the individual galaxies. We note, that black symbols (indicating very low HI ratios) lie predominantly above the diagonal where ram pressure exceeds the gravitational restoring force and are only satellite galaxies.} 
\end{figure} 

In order to calculate the ram pressure on an individual galaxy we use the simple formula for calculating the pressure exerted on a galaxy by the intergalactic medium(IGM). The ram pressure in this case is given by:
\begin{equation}
 P_{ram} = \rho_{IGM} v^{2},
\end{equation}
where $v$ is the relative velocity of the galaxy compared to the medium and $\rho_{IGM}$ is the density of the IGM.
In order to estimate when ram pressure stripping will occur \citet{McCarthy-08} derived a simple condition that, when met, indicated the ram pressure acting on spherically-symmetric gas within a galaxy is efficient:
\begin{equation}
 \rho_{IGM} v^2 > \frac{\pi}{2} \frac{G M_{gal}(R) \rho_{gas}(R)}{R},
\end{equation}
where $M_{gal}(R)$ is the total mass within a radius if R and $\rho_{gas}$ is the the gas density within the same radius. \citet{McCarthy-08} calculate $\rho_{gas}$ as the gas density at radius R rather than within radius R, but we approximate the gas density at R to be the average density of the gas within a sphere. In order to determine if the gas is being stripped from the galaxy we take R to be $5 h^{-1}$ kpc. To estimate the density of the IGM we calculate the average density of a shell between $150 h^{-1}$ kpc and $200 h^{-1}$ kpc. 

We present the results of the ram pressure versus the gravitational restoring force in Figure \ref{fig:carthycond} for both the group and non-group galaxies as well as the central and satellite galaxies in our sample and indicate the HI ratios of the galaxies in question. In the galaxies not residing in groups the ram pressure is generally inefficient with most of the galaxies experiencing very little ram pressure stripping and show high neutral fractions. In the group environment the number of galaxies expected to experience efficient ram pressure stripping increases and the corresponding galaxies show much lower \textsc{Hi} ratios. We observe a similar trend when comparing the central and satellite galaxies in Figure \ref{fig:carthycond}, the satellite galaxies experience much larger ram pressure forces compared to their gravitational restoring force and their \textsc{Hi} are correspondingly lower.

Thus we see that in the group environment ram pressure appears to play a significant role in stripping gas from galaxies and redistributing it through the intergalactic medium, a finding that is in agreement with our earlier observations of extended radial profiles of group galaxies.

\section{Superposition}
\label{sec:Superposition}

In order to address the question of line-of-sight superposition as proposed by \citet{Bordoloi-11} it is worth noting that the method by which we extract the cubes around each galaxy as described in section \ref{subsec:sampleselection} will lead to gas from surrounding galaxies contributing to the radial profiles of the galaxy in question. However we see in figure \ref{fig:StackedCentralvsSat} that the central galaxies, the galaxies we would expect to have the most substructure in their vicinity exhibit sharply dropping profiles when compared to the satellite galaxies.\\

This trend, along with the results presented in figure \ref{fig:carthycond}, in which the satellite galaxies show much less neutral hydrogen associated directly with the galaxy and yet exhibit extended radial profiles suggests that there exists an excess of cold gas in the intergalactic medium group environment. The ram pressure forces illustrated in Figures \ref{fig:carthycond} suggest that this cold gas has been stripped from infalling satellites and is enhancing the radial profiles of the galaxies in a group environment. We conclude that this reservoir of cold gas results in a physical extention of the \textsc{Hi} radial profiles in group environments rather than one due to superposition of neighbouring galaxies.

\section{Conclusions}
\label{secconclusion}

We utilise the mean density region of the \textsc{gimic} suite of simulations to investigate the neutral hydrogen content of galaxies and their surrounding medium. The \textsc{gimic} simulations were developed to investigate the interaction between galaxies and the intergalactic-medium. We calculate the neutral hydrogen content of the gas in the simulation and extract a sample of 488 galaxies to further investigate their \textsc{Hi} properties.\\

We investigate the projected \textsc{Hi} radial profiles of our sample of simulated galaxies in order to observe possible trends resulting from mass and environment. We then stack the radial \textsc{Hi} profiles of the galaxies according to their immediate environment and mass and observe a noticeable difference between the radial \textsc{Hi} profile of those galaxies that reside in groups versus those that reside in a field environment. The galaxies in group environments possess extended radial \textsc{Hi} profiles compared to isolated galaxies. In addition we compare the projected \textsc{Hi} radial profiles of central and satellite galaxies and observe a similar trend in that the satellite galaxies possess extended radial \textsc{Hi} profiles when compared to the central galaxies.\\

In order to investigate the cause for these differences we compare the ram pressure and gravitational restoring forces and estimate the ram pressure experienced by each galaxy. We find that galaxies residing in groups, and satellite galaxies, experience more efficent ram pressure which in turn strips the galaxies of their gas and redistributes it through the intergroup medium. We propose that this could be an important mechanism in extending the radial profiles of individual galaxies.\\

We believe this result is a physical one, due to ram pressure stripping redistributing cold gas into the intergalactic medium, rather than an apparent one, due to superposition of other galaxies in the line of sight, and should be taken into consideration in future surveys particularly in stacking analyses where, in general, environment is not considered.

\section*{Acknowledgements}
The simulations presented here were carried out using the HPCx facility at the Edinburgh Parallel Computing Centre (EPCC)
as part of the EC's DEISA 'Extreme Computing Initiative', and the Cosmology Machine at the Institute for Computational Cosmology of Durham University
We thank the South African Centre for High Performance Computing where some of the simulations and analysis were run, as well as The National Institute of Theoretical Physics, the Square Kilometre Array Project, and the National Research
Foundation for support.  BKG and DC acknowledge the support of
the UK's Science \& Technology Facilities Council (ST/F002432/1 \&
ST/H00260X/1). BKG acknowledges the generous visitor support
provided by Saint Mary's University and Monash University.

\bibliographystyle{mn2e}
\bibliography{ms}

\label{lastpage} 
\end{document}